\documentclass[12pt,a4paper]{article}
\usepackage[utf8]{inputenc}
\usepackage{color}
\usepackage{parskip}
\usepackage{enumitem}
\usepackage{amsmath,amscd}
\usepackage{stmaryrd}
\usepackage{mathrsfs}
\usepackage{amsfonts}
\usepackage{comment}
\usepackage{amssymb,scalerel,stackengine}
\usepackage[margin=1in]{geometry}
\usepackage{graphicx}
\usepackage[leftcaption]{sidecap}
\usepackage{caption}
\usepackage{subcaption}
\usepackage{here}
\usepackage[colorlinks=true,allcolors=blue]{hyperref}
\usepackage{cite}

\newcommand\beq{\begin{equation}}
\newcommand\ee{\end{equation}}
\newcommand\cO{{\cal O}}

\newcommand\dd{\text{d}}

\def\hi{{\hat i}}
\def\hj{{\hat j}}

\def\hr{{\hat r}}

\def\de{\delta}

\def\pd{\partial}

\def\eps{\epsilon}

\def\cM{\mathcal{M}}

\def\cO{\mathcal{O}}

\def\bl{\boldsymbol{\ell}}

\def\th{\theta}

\def\tl{\widetilde}

\newcommand\be[1]{{\boldsymbol{e}_{\hat{#1}}}}

\newcommand\E[2]{{E_{\hat{#1}}^{\;\;#2}}}
\newcommand\barE[2]{{\hat{E}_{\hat{#1}}^{\;\;#2}}}
\def\bu{{\boldsymbol V}}
\def\hmu{{\hat{\mu}}}
\def\hnu{{\hat{\nu}}}

\allowdisplaybreaks

\begin{document}

\begin{flushright}
\hfill{}

\end{flushright}

\vspace{25pt}
\begin{center}
{\Large {\bf Gravitational memory effects and higher derivative actions}}

\vspace{35pt}
{\bf Mahdi Godazgar$^\dagger$, George Macaulay$^\dagger$ and Ali Seraj$^\star$}

\vspace{15pt}

{\it $^\dagger$ School of Mathematical Sciences,
Queen Mary University of London, \\
Mile End Road, E1 4NS, United Kingdom.
\medskip

$^\star$ Centre for Gravitational Waves, Universit\'e Libre de Bruxelles,
International Solvay Institutes, CP 231, B-1050 Brussels, Belgium.}

\vspace{35pt}

\date\today

\vspace{20pt}

\underline{ABSTRACT}
\end{center}

\noindent We show that charges associated with the internal Lorentz symmetries of general relativity, with higher derivative boundary terms included in the action, capture observable gravitational wave effects. In particular, the Gauss-Bonnet charge measures the precession rate of a freely-falling gyroscope, while the Pontryagin charge encodes the relative radial acceleration of freely-falling test masses. This relation highlights the importance of the tetrad formalism and the physical significance of asymptotic internal Lorentz symmetries.



\noindent

\thispagestyle{empty}

\vfill
E-mails: m.godazgar@qmul.ac.uk, g.long@qmul.ac.uk, ali.seraj@ulb.be

\pagebreak

\section{Introduction} \label{sec:intro}

The confluence of a new era of gravitational wave astronomy \cite{PhysRevLett.116.061102} and a new understanding of the relationship \cite{Strom:soft1, Strominger:2014pwa, Pasterski:2015tva, Strom:lec,Nichols:2017rqr,Nichols:2018qac,Seraj:2021qja,Grant:2021hga} between generalized asymptotic BMS charges \cite{bondi,sachs, Barnich:2009se,Campiglia:2014yka,Mao:2018xcw,Godazgar:2020gqd,dualex,Freidel:2021fxf,Strominger:2021lvk}, and gravitational memory effects\cite{Zeldovich:1974gvh,Braginsky:1985vlg,Christodoulou:1991cr,Blanchet:1992br,Favata:2010zu} has led to the tantalising possibility of using observations to constrain semi-classical effects in gravity \cite{Compere:2017wrj, Compere:2019gft, Mao:2020vgh, Seraj:2021qja, Mao:2021eor, Freidel:2021qpz}. An important aspect of this new understanding is the expectation that any gravitational memory effect corresponds to an asymptotic charge generated by a particular asymptotic symmetry.  For example, the standard displacement memory effect \cite{Zeldovich:1974gvh, Braginsky:1985vlg, BT} is related to supertranslation charges \cite{Strominger:2014pwa, Strom:lec}.  Given the plethora of memory effects \cite{Pasterski:2015tva, Zhang:2017rno, Zhang:2017geq, Flanagan:2018yzh, Mao:2018xcw, Hou:2020tnd, Tahura:2020vsa, Tahura:2021hbk} and various extensions of the asymptotic symmetry group \cite{Barnich:2009se, BarnichAspects, Campiglia:2014yka, fakenews, dual0, dualex, Freidel:2021fxf, Guevara:2021abz, Strominger:2021lvk, Geiller:2022vto}, a natural programme to consider is one of relating the various memory effects to the various asymptotic charges, in the hope that one may learn something new about semi-classical/quantum gravity.

Recently, it has been argued \cite{Godazgar:2020gqd} that the most appropriate context in which to consider such questions is the first order tetrad formalism \cite{Barnich:2016rwk, DePaoli:2018erh, Oliveri:2019gvm, Barnich:2019vzx}.  The relevant gravitational action then includes, in principle, any extra terms that do not contribute to the equations of motion, including higher derivative terms \cite{Godazgar:2022foc, Liu:2022uox}. The result is that by applying the covariant phase space formalism \cite{peierls, berg, CWitten, Crknovic88, Lee:1990nz, Iyer:1994ys, WZ} to the asymptotic BMS symmetries of the background, we obtain more charges than may have been expected.  In particular, in addition to the standard BMS charges that have been known in the literature for some time (see e.g.\ \cite{BarTro}), one obtains also dual charges \cite{dual0,dualex}\footnote{An alternative approach which makes the relation between charges and symmetries one to one, is to start from a duality invariant formulation of the theory at hand. In this way, dual charges generate asymptotic symmetries of the dual gauge field \cite{Hosseinzadeh:2018dkh}. Interestingly, the expression of the dual charge is the same in both approaches.}. The contribution of higher derivative terms in the action are not associated with any global charges, such as Bondi mass or angular momentum, or NUT charges.  Nevertheless, the charges derived therefrom are non-trivial \cite{Godazgar:2022foc}.  Moreover, in contrast with the two-derivative terms, there are charges derived from the higher derivative terms that are generated by internal Lorentz transformations.  The status of internal Lorentz transformations is a matter of debate within the literature \cite{Jacobson:2015uqa, Prabhu:2015vua, Barnich:2016lyg, DePaoli:2018erh, Nguyen:2020hot, Godazgar:2020kqd, Freidel:2020xyx, Freidel:2020svx, Freidel:2021fxf, Oliveri:2020xls}.  However, the results of Ref.\ \cite{Godazgar:2022foc} suggest that their physical significance must at least be considered.  In this work, we shall go further by arguing that not only are charges associated with internal Lorentz transformations non-trivial but that they are directly related to physical observables.  

The physical observable that is of particular interest here is the precession of a freely-falling gyroscope in the asymptotic region of spacetime caused by gravitational waves \cite{Seraj:2021rxd,Seraj:2022qyt}, and a net rotation after the passage of gravitational wave dubbed the `gyroscopic memory effect'.  As noted in Refs.\ \cite{Seraj:2021rxd,Seraj:2022qyt}, the expression for the gyroscopic memory effect is clearly related to dual quantities.  However, it is not equal to any of the dual charges previously derived.   

In this paper, we show that the gyroscopic memory effect is related to the dual higher derivative or Gauss-Bonnet charge of \cite{Godazgar:2022foc} that is generated by internal Lorentz transformations.  Furthermore, we show that the Pontryagin charge generated by an internal Lorentz transformation is related to a new subleading memory effect, which we call the radial kick memory.  While, the relation of the precession of a spin vector to charges derived from asymptotic Lorentz transformations is natural, given that the effect is very much frame-dependent, the relation of a radial kick memory effect to an internal Lorentz charge is not immediately obvious.  

One possible explanation for the relation of these memory effects to internal Lorentz charges is that beyond the standard and dual BMS charges derived from the two-derivatve terms in the action, there are no other independent charges and that the internal charges conveniently repackage an expression that may equally be interpreted as a spacetime rotation or even superrotation charge \cite{Barnich:2009se}.  Another possibility is that the internal Lorentz charges are in fact related to various extensions of the BMS group \cite{Campiglia:2014yka, Campiglia:2015yka, Freidel:2021fxf, Guevara:2021abz, Strominger:2021lvk, Geiller:2022vto}.  In order to investigate this, one must first investigate the charges associated with the additional symmetry generators.

Of course, one may choose boundary conditions that freeze out asymptotic internal Lorentz transformations, just as one can choose boundary conditions that freeze out supertranslations and break the BMS group down to the Poincar\'e group.  However, correspondingly, such a choice would amount to a configuration space that precludes a gyroscopic memory effect.  We would argue that this would not be desirable as such a memory effect is a natural physical effect that ought to be reflected in any true configuration space.  Therefore, the boundary conditions ought to be such as to accommodate asymptotic internal Lorentz transformations. This highlights the importance of memory effects as a guide for determining the appropriate boundary conditions that one ought to impose.

This paper is organized as follows. In section \ref{sec:met}, we begin by reviewing the space of solutions that we shall we be working in, namely asymptotically flat spacetimes and recall their asymptotic symmetry generators.  In subsection \ref{sec:frame}, we define a local inertial frame, which is necessary for describing gravitational memory effects, which are treated in subsection \ref{sec:mem}.  In this subsection we introduce the radial kick memory and review briefly the gyroscopic memory effect of Ref.\ \cite{Seraj:2021rxd,Seraj:2022qyt}. In section \ref{sec:charge}, we review the higher derivative charges derived in Ref.\ \cite{Godazgar:2022foc} and make the link between these and the memory effects discussed in section \ref{sec:mem}.  In section \ref{vacuum transition}, we establish the gyroscopic memory effect as a vacuum transition under the residual internal Lorentz symmetry.  We end with some general comments.

\noindent \textbf{Notation:} Latin indices $(a, b,...)$ denote internal Lorentz indices and are raised and lowered with respect to the Lorentz metric $\eta_{ab}$. We use Greek letters $(\mu, \nu,...)$ to denote the spacetime indices.  The vierbein $e^a= {e^a}_{\mu} dx^{\mu}$. The spacetime metric can be expressed in terms of the vierbein as ${ds^2=g_{\mu \nu} dx^\mu dx^\nu= \eta_{ab} e^a e^b}$. Indices $(\hat{\mu}, \hat{\nu},...)$ denote internal Lorentz indices associated with a particular orthonormal frame introduced in section \ref{sec:mem}. Indices $I,J,...$ will denote spacetime indices on the round 2-sphere and will be lowered and raised using the round 2-sphere metric $\gamma_{IJ}$ and its inverse, respectively, except where explicitly stated otherwise.  Finally, we define the curvature 2-form as ${\mathcal{R}_{ab}(\omega)= d\omega_{ab} +\omega_{ac} \wedge {\omega^c}_b}$, where $\omega_{ab}=\omega_{\mu ab}dx^\mu$ is the spin connection. 

\section{Metrics and frames in the far zone} \label{sec:met}
In the absence of cosmological constant, arbitrary localized matter sources lead to asymptotically flat spacetimes, which are most conveniently described in the Bondi gauge
\beq
\label{bog}
g_{rr}=g_{ra}=0,
\qquad
\pd_r\det\left(r^{-2}g_{ab}\right)=0.
\ee
Any such metric is commonly written as 
\beq
\label{e2}
ds^2
=
-e^{2\beta}(F d u^{2}+2 du\,dr)
+r^2h_{IJ}\big(d\th^{I}-\frac{\,U^{I}}{r^2} du\big)\big(d\th^{J}-\frac{\,U^{J}}{r^2}du\big).
\ee
The Bondi gauge is particularly well adapted to null rays generated by the matter source whose center of mass is located at the origin of space $r=0$. The future lightcones of the source are described by $u=const$ surfaces with normal $\bl= e^{-2\beta}\pd_r$, the tangent vector to outgoing affine null geodesics, along which the angles $\th^I$ remain constant. Finally, $r$ is the areal distance, so that the area of coordinate spheres is $4\pi r^2$ irrespective of the metric. In the far zone where $r$ is very large compared to the size of the source, one can solve the Einstein equations perturbatively in the small parameter $1/r$, with the boundary condition that $\lim_{r\to \infty}h_{IJ}(u,r,\th)= q_{IJ}(\th),$ where $q_{IJ}$ is a fixed metric on the round sphere with Ricci scalar $R[q]=2$. Thus we find
\begin{align}
	\label{fex}
	F
	&=
	1-\frac{2 m}{r}+\frac{1}{r^2}\left(\frac{1}{16}C^2
	+\frac{1}{3}D_{I} L^{I}+\frac{1}{4}D_{I} C^{I J}D_{K} C_{J}{}^{K}\right)+\cO(r^{-3}),\\
	\label{bex}
	\beta
	&=
	-\frac{C^2}{32r^2}+\cO(r^{-3}),\\
	\label{gex}
	h_{IJ}
	&=
	q_{IJ}+\frac{1}{r} C_{IJ}+\frac{1}{4r^2} q_{IJ}C^2+\cO(r^{-3}),\\
	\label{uex}
	{U}^{I}
	&=
	-\frac12 D_JC^{IJ}+\frac{1}{r}\left[-\frac{2}{3} L^{I}+\frac{1}{16} D^{I}C^2
	+\frac{1}{2} C^{IJ} D^{K} C_{JK}\right]+\cO(r^{-2}),
\end{align}
where $C_{IJ}$ is the Bondi shear, $D_I$ is the covariant derivative on the  `celestial' sphere $D_Iq_{JK}=0$ and we use the shorthand notation $C^2\equiv C_{IJ}C^{IJ}$. The remaining Einstein equations fix the time derivative of the mass and angular momentum aspects $m$ and $L_I$ in terms of the shear.

The geometric construction of the coordinate system as described above still admits certain residual coordinate transformations given by the BMS generators $x\to x+\xi$ with
\begin{align}
	\xi^{u}=f, \quad \xi^{r}=\frac{r}{2}\left(C^{I} \partial_{I} f-D_{I} \xi^{I}\right), \quad \xi^{I}=Y^{I}-\int_{r}^{\infty} d r^{\prime} \frac{e^{2 \beta}}{r^{\prime 2}}\left(h^{-1}\right)^{I J} \partial_{J} f,
\end{align}
where $f=T\left(x^{I}\right)+\frac{u}{2} D_{I} Y^{I}$ and $Y^I$ is a conformal Killing vector on the sphere. It is well known that these form the asymptotic symmetries of the asymptotic phase space and correspond to well defined charges given by 
\begin{align}\label{BMS}
P_T=\int T m\,,\qquad J_Y=\int Y^I N_I,   
\end{align}
where $N_I$ is the improved angular momentum aspect of Ref.\ \cite{Flanagan:2015pxa}. 

\subsection{Local inertial frames} \label{sec:frame}

An experimenter located in the far zone describes her measurements in a local frame constructed by orthonormal basis vectors $e_{\hmu}{}^\mu$. Assuming that the observer  has velocity $\bu=V^\mu\pd_\mu$, a natural tetrad adapted to the observer and to the radial rays arriving from the source is \cite{Seraj:2021rxd}
\begin{align}\label{eq:SO frame}
\be{0}{}=\bu\,,\qquad
\be{r}=\frac{1}{\gamma}\bl-\bu\,,\qquad 
\be{i}=\frac{\E{i}{I}}{r}\left[\pd_I+\gamma_{IJ}\Big(r^2 v^{J}-U^{J}\Big)\bl\right],
\end{align}
where $\E{i}{I}$ is a zweibein on the sphere constrained by 
\begin{align}\label{zweibein}
h_{I J} \E{i}{I} \E{j}{J}=\delta_{\hat{i} \hat{j}}.    
\end{align}
The condition \eqref{zweibein} still allows for rotations of the transverse vectors by an arbitrary rotation angle $\lambda(u,r,\th^I)$ as $\E{i}{I}\to R_{\hi}{}^\hj\E{j}{I}$. To study physical effects meaningfully, we fix the tetrad further by imposing that transverse directions are induced from a given zweibein $\barE{i}{I}$ on the celestial sphere \cite{Godazgar:2022foc}, i.e.\ 
\begin{align}
    E_{\hi}{}^{I}(u,r,\th^I)=X^{I}{}_{J}(u,r,\th^I) \hat{E}_{\hi}{}^{J}(\th^I)\,,
\end{align}
 where $X_{IJ}=q_{IK}X^K{}_J$ is symmetric and $\barE{i}{I}(\th^J)$ is a time-independent zweibein on the celestial sphere such that 
\begin{equation}
     q_{IJ}\barE{i}{I}\barE{j}{J}=\de_{\hat i\hat j}.
\end{equation}
In physical terms, this corresponds to choosing transverse directions from a specific mapping of the sky through distant stars. The above condition and the asymptotic expansion of $h_{I J}$ imply that
\begin{align}\label{zweibein asymptotic expansion}
	\E{i}{I}=\barE{i}{J}(\de_J{}^I-\frac{1}{2r}C_J{}^I+\frac{1}{16r^2}C^2\de_J{}^I)+\cO(1/r^3).
\end{align}
The above construction fixes all `local Lorentz transformations' except a time-independent rotation in the transverse plane with angle $\lambda(\th^I)$ \cite{Godazgar:2022foc}
\begin{align}\label{rotation}
	\barE{i}{I}\to R_{\hat{i}}{}^{\,\hat{j}}(\lambda)\barE{j}{I}\,,\qquad R_{\hi\hj}=\cos\lambda\, \de_{\hi\hj}+\sin\lambda \,\eps_{\hi\hj}.
\end{align}
It is natural to ask whether this residual transformation corresponds to a nontrivial charge and a physical observable. We will address these questions in the following sections.

The above construction works for general observers with arbitrary velocity. However, to make a concrete statement, we consider an observer that is freely falling and is at rest at some initial time $u=u_0$ before the arrival of GWs. The velocity of the observer is then given by $\bu=\gamma(\pd_u+v^r\pd_r+v^I\pd_I)$ with
\begin{subequations}\label{velocity}
\begin{align}
	\gamma&=1+\frac{m_0}{r}+\frac{\gamma_2}{r^2}\,,\qquad\gamma_2=\int_{u_0}^{u}du' m+\frac{1}{16}\Delta C^2, \\
	v^r&=\frac{\Delta m}{r}-\frac{\gamma_2+\Delta\big(\frac{1}{6}D_I L^I+\frac{1}{8}(D_J C^{IJ})^2\big)}{r^2},\\
	v^I&=-\frac{\Delta D_J C^{IJ}}{2r^2}+\frac{D^I\gamma_2+\Delta\big(-\frac23L^I+\frac12C^{IJ}D^KC_{JK}\big)}{r^3}=U^I+\frac{1}{r^2}D_I \int_{u_0}^{u}du' m,
\end{align}
\end{subequations}
where $\Delta X\equiv X(u)-X(u_0)$.
The observer can use the frame \eqref{eq:SO frame} to construct a Riemann normal coordinate system $X^{\hmu}$ in her lab as follows. For any nearby spacetime event $p$, there is a geodesic which passes through the event and intersects the observer's worldline $\Gamma$ orthogonally at a given proper time $\tau(p)$. The coordinate of the event is given by $X^\hmu(p)=(\tau(p),X^{\hat{m}}(p))$ where $X^{\hat{m}}(p)\be{m}$ is the unique vector orthogonal to $\Gamma$ at $\tau(p)$ whose integral curve reaches the event $p$ at an affine parameter equal to $1$. Here, $\hat{m}=(\hat{r},\hat{i})$. The position $X^\hmu$ of free test particles in the observer's lab obeys the geodesic deviation equation
\begin{equation}\label{geodesic deviation}
	\ddot{X}^{\hmu} = R^{\hmu}{}_{\hat{0}\hat{0}\hnu} \, X^{\hnu} 
\end{equation}
where overdot refers to differentiation with respect to the proper time $\tau$. This is because the length scale of the observer's lab is much smaller than the radius of curvature of the background metric.

\subsection{Persistent gravitational wave effects} \label{sec:mem}
There are several gravitational wave effects that persist after the passage of the wave. We will briefly discuss some of these effects that are relevant for our considerations.

\paragraph{Displacement memory.} One can show that at leading order in $1/r$, the dynamics takes place in the transverse plane for which $R_{\hat{0}\hat{i}\hat{0}\hat{j}}=-\frac{1}{2r}\ddot{C}_{\hat{i}\hat{j}}$, which is nonzero only during the time interval $(\tau_0,\tau_f)$ when the gravitational wave  passes through the observation point. Assuming that the test mass is initially at rest, integrating the geodesic deviation equation \eqref{geodesic deviation} twice over time implies 
\begin{align}\label{displacement memory}
	X^{\hat{i}}(u_f)=X_0^{\hat{i}}+ \frac{1}{2r}\Delta C^{\hat{i}}{}_{\hat{j}}X^{\hat{j}}_0.
\end{align}
where $X_0=X(u_0)$ is the initial position of the test mass. This is the standard displacement memory effect \cite{Zeldovich:1974gvh, Braginsky:1985vlg, BT}. 

\paragraph{Gyroscopic memory.} \label{sec:gyr}
Another type of observable that an observer can measure is the orientation of a freely falling gyroscope in her lab and how it is affected by GWs. This was considered in \cite{Seraj:2021rxd,Seraj:2022qyt}.  Consider a small freely falling gyroscope with proper velocity $V^\mu$, given by equation \eqref{velocity}, and spin $S^\mu$. In the local frame \eqref{eq:SO frame} adapted to the observer outlined above, the rate of precession of the spin vector due to the background geometry is 
\begin{equation}\label{eq:parallel transport}
    \frac{d S^{\hat{\mu}}}{d \tau} \equiv \bu \cdot \nabla S^{\hat{\mu}} = \Omega^{\hat{\mu}}{}_{\hat{\nu}}\, S^{\hat{\nu}}, \qquad  \Omega^{\hat{\mu}\hat{\nu}} \equiv - V^\mu \omega_{\mu}{}^{\hat{\mu}\hat{\nu}}.
\end{equation}
In fact, the $\hat{0}$ component of the spin vector vanishes in this comoving frame and the $\hat{r}$ component is subleading.  Therefore, the pertinent effect is given by
\begin{equation}\label{eq:precession}
     \frac{d S^{\hat{i}}}{d \tau} = \Omega^{\hat{i}}{}_{\hat{j}} S^{\hat{j}}.
\end{equation}
A careful computation gives that\footnote{The precession also contains a trivial term that exists even in the Minkowski background. Cancelling this term corresponds to going to a frame whose spatial basis vectors are tied to distant stars rather that the source. See \cite{Seraj:2021rxd,Seraj:2022qyt} for details.}
\begin{equation} \label{gyrmem}
    \Omega_{\hat{i} \hat{j}} = -\frac{\epsilon_{\hat{i} \hat{j}}}{r^2}\tl\cM\,,\qquad \tl\cM=\frac{1}{8}\left( 2 D_I D_J \widetilde{C}^{IJ} - N_{IJ} \widetilde{C}^{IJ} \right),
\end{equation}
where $\epsilon_{\hat{i} \hat{j}}$ is the alternating tensor with $\epsilon_{\hat{1}\hat{2}} = 1$ and the symmetric trace-free dual tensor \cite{dual0} $\widetilde{C}^{IJ} = \epsilon^{IK} C_{K}{}^{J}$.  The quantity $\tl\cM$, called the dual covariant mass aspect in \cite{Freidel:2021qpz}, which transforms covariantly under asymptotic Diff($S^2$) transformations \cite{Freidel:2021qpz} and constitutes the imaginary part of $\mathring\psi_2$, controls the precession rate of the gyroscope due to non-trivial gravitational effects in the background. In particular, note that in the absence of gravitational radiation and gravito-magnetic monopoles (NUT charges), the right hand side is zero, which implies no precession. For the net  memory effect, one considers the time integral of \eqref{gyrmem} in a sandwich configuration of two non-radiating backgrounds with a burst of gravitational radiation in between \cite{Seraj:2021rxd}, giving a net rotation
\begin{align}\label{net rotation}
  \Delta S^\hi&=\frac{1}{r^2}\,\Delta R^\hi{}_\hj\, S_0^\hj\,,\qquad \Delta R^\hi{}_\hj=\lambda\,\eps^\hi{}_\hj\,,\qquad \lambda=\int\dd u\, \tl\cM.
\end{align}

\paragraph{Radial kick memory.} \label{sec:rad}

There is also a subleading memory effect in the radial direction, which has been neglected in the literature simply because it appears at $O(1/r^3)$. However, this quantity naturally appears in our analysis in the next section. Consider two test masses with the same angular position but with some initial radial separation $X_0^{\hr}$.  Using equation \eqref{geodesic deviation}, we find that to leading order
\begin{equation}
	\ddot X^{\hat{r}} = R_{\hat{0} \hat{r} \hat{0} \hat{r}} X_0^{\hat{r}},
\end{equation}
where the relative acceleration in the $e_{\hat{r}}$ direction due to gravitational effects is given to leading order by
\begin{equation} \label{radmem}
	R_{\hat{0} \hat{r} \hat{0} \hat{r}} = - \frac{2}{r^3} \cM\,,\qquad \cM= m + \tfrac{1}{8} N_{IJ} C^{IJ}
\end{equation}
Note that the first term on the right-hand side is expected from Newtonian gravity, while the second term is a genuine gravitational wave effect. The quantity $\cM$,  called the `covariant mass aspect', constitutes the real part of the second Weyl scalar $\mathring\psi_2$ and transforms covariantly under asymptotic Diff($S^2$) transformations \cite{Freidel:2021qpz}.
Note that as long as $u-u_0$ is small\footnote{To be precise, we require that $u-u_0 \ll r$, which can simply be arranged.} with respect to $r$, we have that the only non-zero component of the deviation vector in the asymptotic orthonormal frame is $s^{\hat{r}} = s^r.$   Considering the time integral of \eqref{radmem} in a sandwich configuration of two non-radiating backgrounds with a burst of gravitational radiation in between, the net radial velocity is 
\begin{align}\label{net rotation0}
	\Delta V^\hr=\frac{1}{r^3}\int\dd u\, \cM.
\end{align}

\section{Higher derivative charges} \label{sec:charge}

In Ref.\ \cite{Godazgar:2020gqd}, it is argued that a systematic classification of asymptotic charges in gravity may be obtained by considering all possible contributions to the gravitational action that do not affect the equations of motion in a first order tetrad formulation. In other words, any combination of terms that gives an Einstein tensor upon variation with respect to the gravitational degrees of freedom.  In particular, dual gravitational charges \cite{dual0, dualex} may be obtained from a Holst term, or Nieh-Yan term in the presence of fermions \cite{Godazgar:2020gqd, Godazgar:2020kqd}.  Following Ref.\ \cite{Godazgar:2020gqd}, charges coming from higher derivative terms in the action, namely the  Gauss-Bonnet and Pontryagin terms 
\begin{align} \label{PGBaction}
I_{GB} = \frac{1}{2}\varepsilon_{abcd} \int_{\mathcal{M}} \mathcal{R}^{ab} \wedge \mathcal{R}^{cd}  \quad \text{and} \quad   I_P &= \frac{i}{2} \int_{\mathcal{M}} \mathcal{R}^{ab} \wedge \mathcal{R}_{ab}
\end{align}
have been studied in Ref.\ \cite{Godazgar:2022foc}.  An important aspect of this study is the role of internal Lorentz symmetries, which has been treated by a number of authors 
\cite{Jacobson:2015uqa, Prabhu:2015vua,Barnich:2016lyg, Barnich:2016rwk, DePaoli:2018erh, Barnich:2019vzx, Oliveri:2019gvm, Godazgar:2020gqd, Godazgar:2020kqd, Nguyen:2020hot, Freidel:2020xyx, Freidel:2020svx, Oliveri:2020xls, Freidel:2021fxf}. In Ref.\ \cite{Jacobson:2015uqa}, the combined action of the diffeomorphisms and internal Lorentz transformations on the vierbein is chosen so that it vanishes for Killing isometries.  However, Ref.\ \cite{Godazgar:2022foc} advocates a different definition for asymptotic symmetries: that the action on the \emph{gauge-fixed} vierbein  must be chosen as to match the action of the diffeomorphisms on the \emph{gauge-fixed} metric.  This means then that the residual internal Lorentz transformations then correspond to the improper internal Lorentz transformations and are, therefore, physically relevant.  Although, the two prescriptions are the same as far as the BMS charges are concerned \cite{Oliveri:2020xls}, there is an important consequence for the existence of asymptotic internal Lorentz generators: the prescription of Ref.\ \cite{Jacobson:2015uqa} freezes them out, while the prescription of Ref.\ \cite{Godazgar:2022foc} allows a functions-worth $\lambda(x^I)$ of such generators, which corresponds to rotations in the two-dimensional $i$ directions.  These generators will then have associated asymptotic charges.  In contrast to the case of two-derivative terms in the action, where the residual internal Lorentz symmetries do not give non-trivial charges \cite{Godazgar:2020kqd}, for higher derivative charges this is no longer the case \cite{Godazgar:2022foc}.  To leading order, the internal Lorentz charges associated with the higher derivative terms in the gravitational action are: 
\begin{align}
\mathcal{Q}^{GB}_{\lambda} &= \frac{1}{r} \int_S  d\Omega \hspace{1mm} \lambda \ \big{(}2D_I D_J \widetilde{C}^{IJ} - N_{IJ}\widetilde{C}^{IJ} \big{)} \label{Lorentz:gb}, \\
\mathcal{Q}^P_{\lambda} &= -\frac{4}{r} \int_S  d\Omega \hspace{1mm}  \lambda \ \big{(} 2 m + \tfrac{1}{4} N_{IJ} C^{IJ} \big{)}. \label{Lorentz:p}
\end{align}

Comparing equations \eqref{gyrmem} and \eqref{Lorentz:gb}, and equations \eqref{radmem} and \eqref{Lorentz:p}, we find a striking resemblance.  In fact, if we choose
\begin{equation}
    \lambda \propto \delta^{(2)} (x-x_0),
\end{equation}
where $x_0$ corresponds to the point on the sphere from which the geodesics emanate, we find an equivalence between the respective expressions.  Hence, the precession rate of a freely-falling gyroscope  corresponds to a Gauss-Bonnet charge density generated by an internal Lorentz symmetry, while the radial acceleration of freely-falling test masses corresponds to a Pontryagin charge density generated by an internal Lorentz symmetry. In the case of the gyroscopic memory effect, the relation to an internal Lorentz symmetry is somewhat intuitive, since the very existence of gyroscopic memory relies on the properties of a frame attached to the gyroscope.  Moreover, we have established an interpretation of this effect as a vacuum transition.  In the case of the radial kick memory, the coincidence of the change in the radial velocity given by equation \eqref{net rotation0} and the charge \eqref{Lorentz:p} is not as clear from a phase space perspective.  Mathematically, this coincidence arises from the fact that $R_{\hat{0} \hat{r} \hat{0} \hat{r}}= R_{\hat{1} \hat{2} \hat{1} \hat{2}}$. 

\section{Gyroscopic memory as vacuum transition}\label{vacuum transition} 
The standard displacement memory can be interpreted as a vacuum transition. Let us briefly review that. In practice, the full spacetime can be considered as a sandwich configuration of two non-radiating backgrounds with a burst of gravitational radiation in between. The two non-radiating regions---also called a \textit{radiative vacuum}---are specified by time-independent shears which take the form $C_{IJ}=D_{\langle I}D_{J\rangle} C$. The function  $C(\th^I)$,  called the supertranslation field, parametrizes the radiative vacuum and has the interesting property that it shifts under a supertranslation $\de_TC=T$. Therefore, the vacua form a homogeneous space under the action of supertranslations. 

A generic radiation process induces a net change in the shear given by 
\begin{align}
\Delta C_{IJ}=C_{IJ}(u\to+\infty)-C_{IJ}(u\to-\infty)=\int du N_{IJ}\,,
\end{align}
which corresponds to a vacuum transition, i.e.\ a shift in the supertranslation field $\Delta C=\de_TC=T$, where $T$ is the induced supertranslation. This symmetry transformation is generated by the action of the corresponding supermomentum $P_T=\int_{S^2}m T$. The physical observable of this vacuum transition is the displacement memory discussed around equation \eqref{displacement memory}.


However, our considerations show that one needs to add another function to specify the  radiative vacuum.  A dyad on the sphere $\barE{i}{J}$ is specified by 
\begin{equation}
    \barE{i}{J}=R_{\hi}{}^{\hj}(\Phi)\bar{E}_{\hj}{}^{J},
\end{equation}
where $R_{\hi}{}^{\hj}(\Phi)$ is a rotation matrix with angle $\Phi$ and $\bar{E}_{\hj}{}^{J}$ is a given reference dyad. Under the symmetry transformation \eqref{rotation}, the vacuum field $\Phi$ is shifted $\de_\lambda \Phi=\lambda$.  The net effect of radiation on asymptotic gyroscopes is a rotation in transverse directions given by equation \eqref{net rotation}. This can be equivalently understood as a rotation in the celestial dyad\footnote{From this point of view, the gyroscopic memory is translated into the `astrometric memory' discussed in Ref. \cite{Madison:2020xhh}.}, which in turn corresponds to a shift in the vacuum field $\Phi$ under the symmetry transformation \eqref{rotation}. 
 A vacuum is thus labeled by the pair $\big(C(\th^I),\Phi(\th^I)\big)$ and a radiation process induces a transition in both of these functions, which correspond to symmetry transformations in the phase space generated by the supermomentum $P_T$ and the Gauss Bonnet charge ${\cal Q}^{GB}_\lambda$  defined respectively in  equations \eqref{BMS} and \eqref{Lorentz:gb}:
\begin{align}
    \{P_T,C_{IJ}\}=D_{\langle I}D_{J\rangle}T\,,\qquad \{{\cal Q}^{GB}_\lambda,	\barE{i}{I}\}=\lambda\eps_{\hat{i}}{}^{\, \hat{j}}	\barE{j}{I}.
\end{align}
The orders at which the memory effect, the charge and the change in the physical field arise are analogous to what happens in the case of the displacement memory effect.  In that case, the memory effect is seen at order $1/r$, which is the order at which the Bondi shear appears in the metric on the 2-space, while the supertranslation charge comes in at leading order.  In the case of the gyroscopic memory effect, the rate of precession is at order $1/r^2$, see equation \eqref{gyrmem}, and the charge is at order $1/r$, see equation \eqref{Lorentz:gb}.  The pertinent physical field here is $\omega_{u}{}^{\hat{i} \hat{j}}$, which also changes at order $1/r^2$; see equation (31) of Ref.\ \cite{Seraj:2021rxd}. This analogy is summarised in the following table:
\begin{center}
\begin{tabular}{ c | c | c | c }
  Memory &Physical observable& Vacuum transition &  Generator\\ 
  \hline
 Displacement  &  $\Delta X^{\hi}=\frac{1}{2r} \Delta C^{\hi}{}_\hj\,X_{0}^{\hat{j}}$ &$\Delta C=\de_TC=T$ & $P_T$ in equation \eqref{BMS} \\  
 \hline
 Gyroscopic   & $\Delta S^\hi=\frac{1}{r^2}\,\Delta R^\hi{}_\hj(\lambda)\, S_0^\hj$& $\Delta \Phi=\de_\lambda\Phi=\lambda$    & ${\cal Q}^{GB}_\lambda$ in equation \eqref{Lorentz:gb}
\end{tabular}
\end{center}
We finish with a comment regarding the observability of such effects.  While there is a good prospect of future gravitational wave detectors observing the displacement memory effect \cite{Favata:2010zu,Lasky:2016knh,Boersma:2020gxx,Islo:2019qht}, the additional memory effects considered in this paper are subleading effects, making them more difficult to observe.  Nevertheless, the fact that we have related potentially observable memory effects to charges derived from higher derivative terms in the gravitational action, offers the possibility, in principle, of constraining observationally higher derivative terms, which are important in semi-classical gravity. 
\section{Discussion}
In this paper, we have extended the known examples of a correspondence between asymptotic symmetries and memory effects to two further cases.  As to whether there must in general be a relation between these two physical quantities, we would expect this to be the case, even though there is no proof yet. The reason that we expect this statement to be true is the following:	A memory effect is typically understood in terms of a transition between two non-radiative spacetimes. However, a non-radiative configuration (which we may call a radiative vacuum, i.e.\ a no graviton state) is also characterized by charges. This point is investigated in detail recently in \cite{Compere:2022lzx}. 

In this work, we have treated the Gauss Bonnet and Pontryagin terms as independent, namely, we assume that they come with independent and arbitrary coupling constants in the action that we can tune. Indeed, what we call the Gauss Bonnet and Pontryagin charges are derived by varying the charge corresponding to the residual internal Lorentz symmetry with respect to these coupling constants. Similarly supermomenta and dual supermomenta correspond to the real and imaginary parts of the supertranslation charge.  However, another perhaps more satisfactory approach is that the symmetries and charges of the theory are derived after the action has been fixed once and for all. In order to do this one has to implement a ``duality invariant formulation'' of the theory.  While such a formulation has been worked out in the case of Maxwell theory \cite{Hosseinzadeh:2018dkh}, whether this can be done for non-linear gravity is an open problem.

\section*{Acknowledgements}
We would like to thank Guillaume Bossard, Sangmin Choi and Andrea Puhm for useful discussions.
M.G.\ is supported by a Royal Society University Research Fellowship; G.M.\ is supported by a Royal Society Enhancement Award and A.S. is partially supported by the
ERC Advanced Grant ``High-Spin-Grav", and the ERC Starting Grant ``Holography for realistic black holes''. 

\bibliographystyle{utphys}
\bibliography{mem}

\end{document}